\documentclass[prd,twocolumn,showpacs,floatfix,amsmath,amssymb,floatfix]{revtex4}
\usepackage{graphicx,color,dcolumn,booktabs,bm,mathrsfs}
\usepackage{longtable,lscape}
\usepackage{txfonts}
\usepackage{overpic}
\usepackage{amssymb}
\usepackage{indentfirst}
\usepackage{feynmf}   
\usepackage{slashed}  
\usepackage{cases}
\usepackage{color}
\usepackage{multirow}
\usepackage{epstopdf}
\usepackage{graphicx,color,dcolumn,booktabs,bm}

\usepackage[colorlinks,
            citecolor=blue,
            anchorcolor=red,
            menucolor=red,
            linkcolor=red,
            filecolor=red,
            runcolor=red,
            urlcolor=blue]{hyperref}
\begin{document}
\def\pca{P_c^+(4380)}
\def\pcb{P_c^+(4450)}
\def\pcc{P_c^+(4312)}
\def\pcd{P_c^+(4440)}
\def\pce{P_c^+(4457)}
\title{Double Exotic States Productions in Pion and Kaon Induced Reactions}
\author{Jing Liu$^1$}\email{ jingliu@seu.edu.cn}
\author{Dian-Yong Chen $^{1,3}$ \footnote{Corresponding author}} \email{chendy@seu.edu.cn}
\author{Jun He$^{2,3}$} \email{junhe@njnu.edu.cn}
\affiliation{
$^1$ School of Physics, Southeast University,  Nanjing 210094, China\\
$^2$Department of Physics and Institute of Theoretical Physics, Nanjing Normal University, Nanjing 210097, China\\
$^3$Lanzhou Center for Theoretical Physics, Lanzhou University, Lanzhou 730000, China
}
\begin{abstract}
In the present work, we investigate the production of $P_{c}$ and $Z_c(3900)/Z_{cs}(3985)$ states in the processes of $\pi p \rightarrow Z_{c}(3900) P_{c} $/$Kp\rightarrow  Z_{cs}(3985) P_{c}$ with an effective Lagrangian approach. Our estimations indicates that the cross sections depends on the model parameters and beam energies, in particular, the cross sections for $\pi p \to Z_c(3900) P_c(4312)$ and $\pi p \to Z_c(3900) P_c(4440)$ are similar and can reach up to 10 nb, while those for $\pi p \to Z_c(3900) P_c(4457)$ can be 30 nb. As for $Kp \to Z_{cs}(3985)P_c$, the cross sections are about 2.5 times smaller than those for $\pi p \to Z_c(3900) P_c$. The discussed processes in the present work, especially $\pi p \to Z_c(3900) P_c(4457)$, may be accessible by the the high-energy pion and kaon beams in the facilities of J-PARC and COMPASS.
\end{abstract}

\maketitle
\section{Introduction}
\label{sec:introduction}

The investigations of exotic states beyond the conventional hadrons are one of the most intriguing topics in hadron physics in the past two decades. Since the observations of $X(3872)$ in 2003~\cite{Choi:2003ue}, a series of new hadron states have been observed (see Refs~\cite{Klempt:2007cp,Brambilla:2010cs,Chen:2016qju,Lebed:2016hpi,  Guo:2017jvc,Esposito:2016noz,Ali:2017jda, Liu:2019zoy,Brambilla:2019esw,Dong:2017gaw} for recent reviews of experimental and theoretical status), which turns a new page of searching exotic states. Among these new hadron states, some of them are particular interesting, such as the series of $Z_c/Z_{cs}$~\cite{Ablikim:2013mio,Liu:2013dau, Ablikim:2013xfr, Collaboration:2017njt, Ablikim:2013wzq, Ablikim:2013emm, Ablikim:2015tbp} and $P_c/P_{cs}$ states~\cite{Aaij:2015tga, Aaij:2019vzc, Aaij:2020gdg, Ablikim:2014dxl}, which are good candidates of~\cite{Liu:2019zoy, Chen:2020yvq}and pentaquark states~\cite{Ozdem:2021ugy,Wang:2020eep,Chen:2016ryt} respectively.

The tetraquark candidate $Z_c(3900)$ was first observed by the BES III and Belle Collaboration in 2011~\cite{Ablikim:2013mio, Liu:2013dau} in the $J/\psi \pi^\pm$ invariant mass distributions of $e^+ e^- \to \pi^+ \pi^- J/\psi$. Later the CLEO-c confirmed the existence of $Z_c^\pm(3900)$ by using the data of $e^+ e^-\to \pi^+ \pi^- J/\psi$ at $\sqrt{s}=4170 $ MeV~\cite{Xiao:2013iha}. Moreover, the CLEO-c Collaboration also reported the neutral isospin partner of $Z_c^\pm(3900)$, i.e., $Z_c^0(3900)$ for the first time ~\cite{Xiao:2013iha}, which was further observed by the BES III Collaboration ~\cite{Ablikim:2015tbp}. Similar to $Z_c(3900)$, the BES III Collaboration reported another tetraquark candidate $Z_c(4020)$ in the $h_c \pi $ invariant mass spectrum of $e^+ e^- \to \pi^+ \pi^- h_c$~\cite{Ablikim:2013wzq} and $e^+ e^- \to \pi^0 \pi^0 h_c$~\cite{Ablikim:2014dxl}. Besides the hidden charm channel, these two tetraquark candidates have also been observed in the open charm channels~\cite{Ablikim:2013xfr, Ablikim:2013emm}. In the $D^\ast \bar{D}$ invariant mass spectrum of $e^+ e^- \to (D^\ast \bar{D})^\pm \pi^\mp$, a state named $Z_c(3885)$ was reported by the BES III Collaboration in 2013 ~\cite{Ablikim:2013xfr}, while in the $D^\ast \bar{D}^\ast $ invariant mass spectrum of $e^+e^- \to D^\ast \bar{D}^\ast \pi$, one state, $Z_c(4015)$ was also reported by the BES III Collaboration in the same year~\cite{Ablikim:2013emm}. As the strange partner of $Z_c(3900)$, $Z_{cs}(3985)$ was observed in the $D_s^- D^{\ast 0} +D_s^{\ast -} D^0$ invariant mass spectrum of $e^+ e^- \to K^+ (D_s^- D^{\ast 0} +D_s^{\ast -} D^0)$ by BES III Collaboration~\cite{Ablikim:2020hsk}, which makes  the charmonium-like tetraquark family abundant. More recently, the LHCb collaboration reported a structure, $Z_{cs}(4003)$, in the $J/\psi \phi$ invariant mass spectrum of $B^+ \to J/\psi \phi K^+$ \cite{Aaij:2021ivw}, the mass of $Z_{cs}(4003)$ is consistent with the one of $Z_{cs}(3985)$, but the width are much different. The measured resonances parameters of these tetraquark candidates are collected in Table \ref{Tab:Exp}. For comparison, we also present the thresholds near the observed masses of these tetraquark candidates.

As pentaquark candidates, $P_c(4380)$ and $P_c(4450)$ were firstly observed in the $J/\psi p$ invariant mass spectrum of $\Lambda_b\to K J/\psi p$ decay process by the LHCb Collaboration in 2015~\cite{Aaij:2015tga}. The lower mass state is a broader one with $\Gamma=205 \pm 18 \pm 86$ MeV, while the higher mass state is much narrow with the width to be $39 \pm 5 \pm 19$ MeV~\cite{Aaij:2015tga}. In 2019, the LHCb Collaboration reanalysized the same process with a data sample to be nine times more than the one used in Ref.~\cite{Aaij:2015tga} and reported a new narrow pentaquark state $P_c(4312)$~\cite{Aaij:2019vzc}, and the $P_c(4450)$ pentaquark structure was observed to consist of two narrow overlapping peaks, $P_c(4440)$ and $P_c(4457)$~\cite{Aaij:2019vzc}. Later in 2020, the LHCb Collaboration reported a open-strange and hidden-charm pentaquark state, named $P_{cs}^0(4459)$ in $J/\psi \Lambda$ invariant mass spectrum of $\Xi_b^- \to K^- J/\psi \Lambda$ decay process~\cite{Aaij:2020gdg}. The resonance parameters of these pentaquark states are also collected in Table~\ref{Tab:Exp} .

\renewcommand\arraystretch{1.5}
\begin{table*}[t]
 \centering
 \caption{A summary the experimental information of $Z_c/Z_{cs}$ and
 $P_{c}/P_{cs}$ states.\label{Tab:Exp}}
 \begin{tabular}{p{1.2cm}<\centering  p{2cm}<\centering p{3cm}<\centering p{3cm}<\centering p{4cm}<\centering p{2cm}<\centering}
 \toprule[1.2pt]
 State  & Threshold & Mass (MeV) & Width (MeV) & Channel & Experiment\\
 \midrule[1.2pt]
 $Z_c^\pm(3900)$ & $D^\ast\bar{D} $ & $3899.0 \pm 3.6\pm 4.9$  & $46 \pm
 10 \pm 20$  & $e^+ e^- \to \pi^+ \pi^- J/\psi$
  & BES III~\cite{Ablikim:2013mio} \\
   &   & $3894.5 \pm 6.6\pm 4.5$  & $63 \pm
 24 \pm 26$  &
  &  Belle~\cite{Liu:2013dau} \\
     &  & $3901 \pm 4 $  & $58 \pm
 27 $  &
  & CLEO-c~\cite{Xiao:2013iha} \\
 $Z_c^0(3900)$   &  & $3886 \pm 4 \pm 2$  & $37 \pm
 4 \pm 8 $  & $e^+ e^- \to \pi^0 \pi^0 J/\psi$
  & CLEO-c~\cite{Xiao:2013iha} \\
   &  & $3894.8\pm 2.3\pm 3.2$  & $29.6\pm 8.2\pm 8.2$  &
  &  BES III~\cite{Ablikim:2015tbp}\\
$Z_c^\pm(3885)$ &   & $3883.9 \pm 1.5\pm 4.2$  & $24.8 \pm
 3.3 \pm 11.0$  & $e^+ e^- \to (D^{*}\bar{D})^{\pm}\pi^{\mp}$
  & BES III~\cite{Ablikim:2013xfr} \\
\midrule[1.2pt]

 $Z_c^\pm (4020)$ & $D^{*}\bar{D}^{*}$ & $4022.9\pm 0.8\pm 2.7$  &
 $7.9\pm 2.7\pm 2.6 $ & $e^+ e^- \to \pi^+ \pi^- h_{c}$
 & BES III~\cite{Ablikim:2013wzq}\\
$Z_c^0 (4020)$  &   & $4023.9\pm2.2\pm3.8$& $--- $ &$e^+ e^- \to \pi^0
 \pi^0 h_{c}$ &BES III~\cite{Ablikim:2014dxl}\\
 $Z_c^\pm (4025)$ &   & $4026.3 \pm 2.6\pm 3.7$  & $24.8
 \pm 5.6\pm 7.7$  & $e^+ e^- \to (D^{*}\bar{D}^{*})^\pm \pi^\mp$
  & BES III~\cite{Ablikim:2013emm} \\
 \midrule[1.2pt]

$Z_{cs}(3985)$ & $D_{s}^- D^{*0}/D_{s}^{*-} D^{0} $ &
$3982.5_{-2.6}^{+1.8}\pm2.1$  & $12.8_{-4.4}^{+5.3}\pm3.0 $  & $e^+ e^-
\to K^{+}(D^{-}_{s}D^{*0}+D^{*-}_{s}D^{0})$
  & BES III~\cite{Ablikim:2020hsk} \\
$Z_{cs}(4003)$ & $ $ & $4003\pm6_{-14}^{+4}$  & $131\pm15\pm26$
 & $B^{+}\rightarrow J/\psi \phi K^{-}$
& LHCb~\cite{Aaij:2021ivw}
\\
 \midrule[1pt]
 $ \pca$  &$\Sigma^*_{c}\bar{D}$& $4380\pm8\pm9$& $205\pm18\pm86$
 &$\Lambda^{0}_{b}\rightarrow J/\psi K^{-}p$  &
 LHCb~\cite{Aaij:2015tga}\\
 $ \pcb$  &$\Sigma_c\bar{D}^*$&
 $4457.3\pm0.6_{-1.7}^{+4.1}$ &$6.4\pm2.0_{-1.9}^{+5.7}$ &  &LHCb ~\cite{Aaij:2015tga}\\
 $\pcc$&$\Sigma_{c}\bar{D}$& $4311.9\pm 0.7_{-0.6}^{+6.8}$ &
 $9.8\pm2.7_{-4.5}^{+3.7}$  &
 &LHCb~\cite{Aaij:2019vzc} \\
 $\pcd$ &$\Sigma_{c}\bar{D}^{*} $ &$ 4440.3\pm 1.3_{-4.7}^{+4.1}$&
 $20.6\pm4.9_{-10.1}^{+8.7}$  &  &LHCb ~\cite{Aaij:2019vzc}\\
 $\pce$ &$\Sigma_{c}\bar{D}^{*}$& $4457.3\pm 0.6_{-1.7}^{+4.1}$&
 $6.4\pm2.0_{-1.9}^{+5.7}$ &
 &LHCb~\cite{Aaij:2019vzc}\\
 \midrule[1pt]
 $P^0_{cs}(4459)$ &$\Xi_{c}\bar{D}^{*}$& $4458.8\pm2.9_{-1.1}^{+4.7}$ &
 $17.3\pm6.5_{-5.7}^{+8.0}$ &$\Xi^{-}_{b}\rightarrow J/\psi K^{-}\Lambda$
 & LHCb~\cite{Aaij:2020gdg}\\
 \bottomrule[1.2pt]
 \end{tabular}
 \end{table*}

 These observations stimulated theorists great interests to investigate the nature and internal structure of these tetraquark and pentaquark states. The observed channels indicates that the most possible quark components of $Z_c/Z_{cs}$ states should be $c\bar{c} q\bar{q}/c\bar{c} q \bar{s}$, while those of $P_c/P_{cs}$ should be $c\bar{c} uud/c\bar{c} uds$, which means that these states can be naturally considered as compact multiquark states. In the compact tetraquark scenario, the mass spectrum~\cite{Faccini:2013lda,Braaten:2013boa,Goerke:2016hxf,Qiao:2013raa,Wang:2013vex,Deng:2014gqa,Agaev:2016dev,Qiao:2013dda,Wang:2013exa,Giron:2021sla,Shi:2021jyr, Liu:2021xje} and decay properties \cite{Agaev:2016dev,Dias:2013xfa,Wang:2017lot,Esposito:2014hsa, Guo:2020vmu} of $Z_c^{(\prime)}/Z_{cs}$ have been investigate by various methods, such as QCD sum rule, constituent quark model. The estimations by QCD sum rule~\cite{Chen:2015moa,Wang:2015epa,Wang:2019got} and constituent quark model \cite{Ortega:2016syt,Park:2017jbn,Weng:2019ynv,Zhu:2019iwm,Giron:2021sla} supported the compact pentaquark interpretations of $P_c/P_{cs}$.

  As presented in Table \ref{Tab:Exp}, all these observed tetraquark and pentaquark candidates are close to the thresholds of hadron pairs. Then, all these states can be considered as molecular states. In the $D^\ast \bar{D}^{(\ast)}/D_s \bar{D}^\ast$ molecular scenario, the mass spectrum of $Z_c^{(\prime)}/Z_{cs}$ were estimated by using QCD sum rule ~\cite{Chen:2015ata,Qiao:2013dda,Wang:2015epa,Wang:2019got} and potential model with one-boson-exchange potential~\cite{Liu:2009qhy,Liu:2008fh,Sun:2012zzd,He:2013nwa}.   The decay ~\cite{Chen:2015igx,Gutsche:2014zda} and production properties ~\cite{Guo:2019fdo,Chen:2016byt,Lin:2013mka,Huang:2015xud,D0:2018wyb,Belle:2003nnu,Belle:2008fma} were also investigated in an effective Lagrangian approach. As for $P_c/P_{cs}$, their properties, including mass spectrum ~\cite{Chen:2019bip,Chen:2019asm,He:2019ify,Liu:2019tjn} decay properties ~\cite{Cheng:2019obk,Guo:2019kdc,Xiao:2019aya,Guo:2019fdo} and production behaviors ~\cite{Wang:2019krd,Wu:2019rog,Wang:2019dsi} have been investigated in the molecular scenarios. Besides the molecular interpretation, these tetraquark and pentaquark states were also considered as structures generated by some special kinematics mechanisms, such as cusp effect~\cite{Liu:2013vfa,Swanson:2014tra}, initial single chiral partial emission mechanism ~\cite{Chen:2011xk,Chen:2013coa}, and triangle singularity ~\cite{Shen:2020gpw}.

As one of important aspect of evaluating the properties of the tentraquark and pentaquark candidates, the productions of these states have been widely investigated. For example, in Ref.~\cite{Chen:2016byt}, the production process $Y(4260) \to Z_c(3900) \pi$ was investigated in a molecular scenario, where $Y(4260)$ and $Z_c(3900)$ were considered as $D \bar{D}_1(2420)+h.c$ and $D\bar{D}^\ast +h.c$ molecular states, respectively. The production of $Z_c^{(\prime)}$ states from $B_c$ decays were predicted in Ref. \cite{Wu:2019vbk}. Moreover, the productions of $Z_c$ in the  $\pi p$~\cite{Huang:2015xud}, $\gamma p$~\cite{Deng:2014gqa} process have also been investigated in an effective Lagrangian approach. As for $P_{c}/P_{cs}$ states, the observed production process has been investigated in Ref. \cite{Wu:2021dmq, Wu:2019rog} and the production of these pentaquark states from $\pi p$~\cite{Wang:2019dsi} scattering and bottom baryon decay processes~\cite{Feijoo:2015kts,Lu:2016roh,Chen:2015sxa,Shen:2020gpw}. It worth to mention that in the high energy pion/kaon induced reaction, the contributions from $s$- and $u$-channels are much smaller than the $t$-channel contribution, which indicates that the $s$- and $u$-channels can usually be ignored. Considering the strong coupling between $Z_c$ and $J/\psi \pi $, the process $\pi p\to Z_c p$ could occur via a $J/\psi $ exchange while the $J/\psi$ couples to the proton via vector meson dominance as indicated in Ref.~\cite{Huang:2015xud}. The predicted cross section are on the order of magnitude of 1 nb when considering $J/\psi \pi$ channel to be the dominant one of $Z_c$. Actually, the exchanged $J/\psi$ and proton target can strongly couples to the pentaquark candidate $P_c$ states directly. Thus, one can construct a possible double exotic production process, i.e., $\pi p \to Z_c P_c$. In a same way, one can  replace the pion beam by the kaon beam to construct a very similar double exotic production process $K p \to Z_{cs} P_c$. In the present work, we adopt an effective Lagrangian approach to estimate the cross sections for these two kinds of double exotic production processes, which may be accessible by the the high-energy pion and kaon beams in the facilities of J-PARC~\cite{Austregesilo:2018mno, Kumano:2015gna} and COMPASS~\cite{Nerling:2012er}.

 This work is organized as follows. After the introduction, we present the effective Lagrangian s used in the present work and the amplitudes in Section~\ref{sec:Lagrangians}. The numerical results of the total cross section, the differential cross sections and cross section ratios are present in Section~\ref{sec:Numerical} and we give a brief summary of this work in the last section.

\section{Pion and Kaon induced production on a proton target}
\label{sec:Lagrangians}
\begin{figure}[htb]
\begin{tabular}{ccc}
\centering
\includegraphics[width=8.5cm]{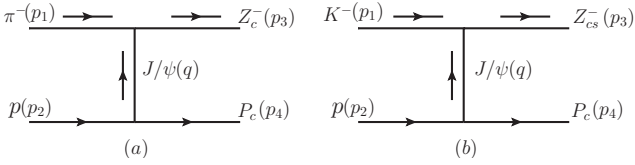}&
 \end{tabular}
\caption{A sketch diagram contributing to
$\pi^{-} p\rightarrow Z_{c}(3900)P_{c}$ (diagrams (a)) and $K^{-} p\rightarrow Z_{cs}(3985)P_{c}$(diagrams (b)), where $P_c $ can be $P_c(4312)$, $P_c(4440)$ and $P_c(4457)$, respectively.\label{Fig:Pc}}
\end{figure}

As indicated in Fig.~\ref{Fig:Pc}, the double exotic production process $\pi p \to P_c Z_c$ occur via a $J/\psi$ exchange. These diagrams are estimated in an effective Lagrangian approach, where the relevant effective Lagrangians read~\cite{Wang:2019krd,Wang:2015jsa},
\begin{eqnarray}
\mathcal{L}_{\pi\psi Z_{c}}&=&\frac{g_{\pi\psi Z_{c}}}{m_{Z_{c}}}(\partial_{\mu }\psi_{\nu}\partial^{\mu}\pi Z_{c}^{\nu}-\partial_{\mu}\psi_{\nu}\partial^{\nu}\pi Z_{c}^{\mu})\nonumber\\
\mathcal{L}^{1/2^{-}}_{P_{c}\psi p}&=&g^{1/2^{-}}_{P_{c}\psi p}\bar{p}\gamma_{5}\gamma_{\mu}P_{c}\psi^{\mu}+H.c.\nonumber\\
\mathcal{L}_{P_{c}\psi p}^{3/2^{-}}&=&\frac{-ig_{P_{c}\psi p}^{3/2^{-}}}{2m_{p}}\bar{P}_{c\mu}\psi^{\mu\nu} \gamma_{\nu} p +H.c. \label{Eq:Lag}
\end{eqnarray}
where the $\psi^{\mu\nu}=\partial^{\mu} \psi^{\nu}-\partial^{\nu} \psi^{\mu}$. The $\pi$, $p$, $P_{c}$, $\psi$ and $Z_{c}(3900)$ are the
pion meson, the proton, the $P_{c}$ state, $J/\psi$ meson and $Z_{c}(3900)$ meson fields, respectively. As for the $K \psi Z_{cs}$, the effective Lagrangian is the same in the form as the one of $\pi \psi Z_c$. The concrete values of the relevant coupling constants will be discussed in the next section.

With the above effective Lagrangians, one can obtain the amplitudes corresponding to the diagrams in Fig.~\ref{Fig:Pc}, which are,
\begin{eqnarray}
\mathcal{M}^{1/2^{-}}&=&g^{1/2^{-}}_{P_{c}p\psi}[\bar{u}(p_{4},m_{4})\gamma_{\mu}\gamma_{5}u(p_{2},m_{2})]\nonumber\\
        &\times&g_{Z_{c}\pi\psi}[(-iq^{\rho})(-p^{\rho}_{1})g^{\nu\beta}-(-q^{\rho})(-p^{\nu}_{1})
         g^{\rho\beta}]\epsilon^{\beta}_{3}\nonumber\\
        &\times & \frac{-g^{\mu\nu}+q^{\mu}q^{\nu}/m^{2}_{\psi}}{q^{2}-m^{2}_{\psi}}
        F(q^2,m_{\psi})\\
\mathcal{M}^{3/2^{-}}&=&\frac{g^{3/2^{-}}_{P_{c}p\psi}}{2m_2}[\bar{u}_{\mu}(p_{4},m_{4})(-iq^{\mu})g^{\nu\eta}
             -(-iq^{\nu})g^{\mu\eta}
             \gamma_{\nu}u(p_{2},m_{2})]\nonumber\\ &\times&
             g_{Z_{c}\pi\psi}[(-iq^{\rho})(-ip^{\rho}_{1})g^{\sigma\xi}-(-iq^{\rho})
             g^{\rho\xi}(-ip^{\sigma}_{1})] \epsilon^{\xi}_{3}\nonumber\\ &\times&
            \frac{-g^{\eta\sigma}+q^{\eta}q^{\sigma}/m_{\psi}^{2}}{q^{2}-m_{\psi}^{2}}
             F(q^2,m_{\psi})
\end{eqnarray}
where $\mathcal{M}^{1/2^{-}}$ and $\mathcal{M}^{3/2^{-}}$ are the amplitudes of $\pi p \to Z_c P_c$ with the $J^P$ quantum numbers of $P_c$ state are $\frac{1}{2}^-$ and $\frac{3}{2}^-$, respectively. In particular, $\mathcal{M}^{1/2^-}$ can be the amplitudes of $\pi p\to Z_c P_c(4312)$ and $\pi p\to Z_c P_c(4440)$, while the amplitude $\mathcal{M}^{3/2^-}$ corresponds to the process $\pi p \to Z_c P_c(4457)$.  In the amplitudes, a form factor, $F(q^{2},m_{\psi})$ is introduced to depict the internal structure and the off-shell effects of the exchanged $J/\psi$. In this work, we adopt the form factor in the monopole form, which is ~\cite{Chen:2013cpa,Chen:2014ccr},
\begin{eqnarray}
F(q^{2},m_{\psi})=\frac{m^{2}_{\psi}-\Lambda^{2}}{q^{2}-\Lambda^{2}}.
\end{eqnarray}
Here, $\Lambda$ is the cutoff parameter and of order of 1 GeV, $q_{\psi}$ and $m_{\psi}$ are the momentum and the mass of the exchanged $J/\psi$ meson, respectively.

 With the above amplitudes, one can obtain the differential cross sections of $\pi p \to Z_c P_c$, which reads,
 \begin{eqnarray}
 \frac{d\sigma}{dcos\theta}=\frac{1}{32 \pi s}\frac{|\vec{p}_{f}|}{|\vec{p}_{i}|}(\frac{1}{2}\overline{|\mathcal{M}|^{2}}),
 \end{eqnarray}
 where $s=(p_{1}+p_{2})^{2}$ is the center mass energy and $\theta$ is the angle of the outgoing $P_{c}$ states related to the pion beam, while the $\vec{p}_{i}$ and $\vec{p}_{f}$ are the momentum of initial pion beam and final $P_{c}$ states in the center of mass frame, respectively.

 \section{Numerical results and discussion}
\label{sec:Numerical}

\renewcommand\arraystretch{1.35}
\begin{table}[h]
 \centering
 \caption{The values of the coupling constants estimated by the partial widths of the corresponding process.\label{Tab:CP}}.
 \begin{tabular}{p{2cm}<\centering p{2cm}<\centering p{2cm}<\centering p{2cm}<\centering }
 \toprule[1 pt]
 Coupling  & $g^{1/2}_{P_c(4312) \psi p}$ & $g^{1/2}_{P_c(4440) \psi p}$ & $g^{3/2}_{P_c(4457) \psi p}$ \\
 Value &$0.11$ &$0.14$ & $0.08$ \\
 \midrule[1pt]
 Coupling  & $g_{Z_c \psi \pi}$ & $g_{Z_{cs} \psi K}$ &  $ $\\
 Value & $0.562$ & $0.364$ & $ $  \\
  \bottomrule[1 pt]
 \end{tabular}
 \end{table}

 \begin{figure*}[t]
\includegraphics[width=16 cm]{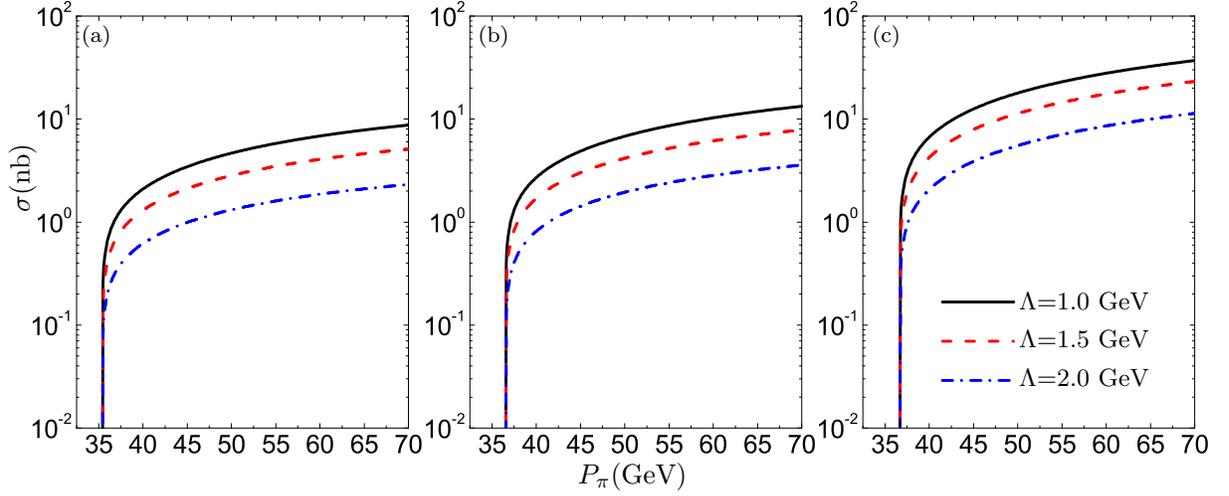}
 \caption{The cross sections for $\pi p \to Z_c(3900) P_c$ depending on the momentum of pion beam. Diagrams(a),(b) and (c) correspond to cross sections for $P_{c}(4312)$ $P_{c}(4440)$ and $P_{c}(4457)$ production, respectively. }\label{Fig:CS-pip}
\end{figure*}

\begin{figure*}[htb]
\includegraphics[width=16cm]{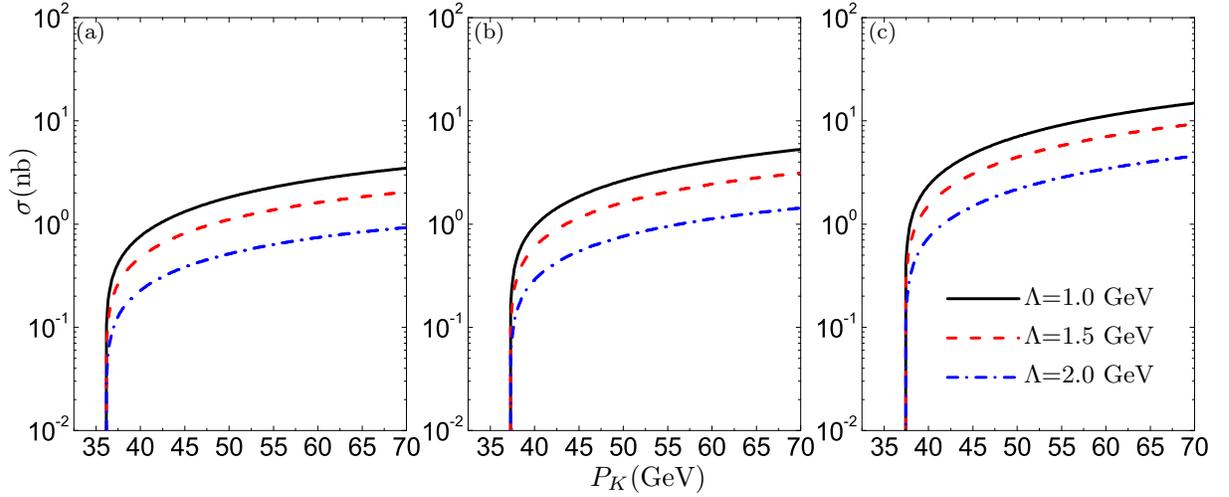}
 \caption{The same as Fig.~\ref{Fig:CS-pip} but for $Kp\to Z_{cs}(3985) P_c$. }\label{Fig:CS-Kp}
\end{figure*}

\subsection{Coupling Constants}
In the present work, an effective Lagrangian approach is adopt to estimate the cross sections for the double exotic states production process. The effective coupling constants in Eq.~(\ref{Eq:Lag}) can be determined by using the corresponding partial widths.  Form the effective Lagrangians listed in Eq.~(\ref{Eq:Lag}), the partial widths of $P_c(p_0) \to J/\psi(p_1) p (p_2)$ and $Z_c(p_0) \to J/\psi(p_1) \pi(p_2) $,
 \begin{eqnarray}
 \Gamma_{P_{c}\rightarrow J/\psi p}^{1/2^{-}}
            &=&g^{2}_{P_{c}J/\psi p}\frac{|\vec{p}|}{8\pi m_{0}^{2}m_{1}^{2}}[(m_2+m_0)^2-m_1^2]\nonumber \\
            &\times&[(m_{2}-m_{0})^{2}+2m_{1}^{2}]\nonumber \\
 \Gamma_{P_{c}\rightarrow J/\psi p}^{3/2^{-}}
            &=&g_{P_{c} J/\psi p}^{2}\frac{|\vec{p}|}{192\pi m_0^4m^2_2}[(3m_0^6-m_0^4(m_1^2+5m^2_2)\nonumber \\
            &+&12m_0^3m_1^2m_2+m_0^2(m_2^4-m_1^4)-(m_1^2-m_2^2)^3 ]\nonumber \\
 \Gamma_{Z_c \to J/\psi \pi }&=&
      g_{Z_cJ/\psi \pi}^{2}\frac{|\vec{p}|}{96 \pi m_0^4}[2m_0^6-m_0^4(3m_1^2+4m_2^2)\nonumber \\
              &+&2m_0^2(3m_1^2m_2^2+m_2^4)+(m_1^3-m_1m_2^2)^2].
           \label{Eq:PW}
 \end{eqnarray}
with  $|\vec{p}|=\lambda(m^{2}_{0},m_1^{2},m^{2}_{2})/2m_{0}$ and $\lambda$ to be the K\"{a}llen function with the definition $\lambda(x,y,z)\equiv \sqrt{(x^{2}+y^{2}+z^{2}-2xy-2xz-2yz}$.

As for $Z_c(3900)$, the partial width ratio of $D^\ast \bar{D}$ and $J/\psi \pi$ channels is measured to be~\cite{Ablikim:2013xfr},
\begin{eqnarray}
 \frac{\Gamma(Z_{c}\rightarrow \bar{D}^{*}D)}{\Gamma(Z_{c}\rightarrow J/\psi \pi)}=6.2\pm1.1\pm2.7. \label{Eq:BRZc}
 \end{eqnarray}
With the assumption that $Z_c(3900)$ dominantly decay into $D^\ast \bar{D}$ and $J/\psi \pi$, one can estimate the partial width of $Z_c(3900) \to J/\psi \pi$ with the measured width, and then the coupling constants $g_{Z_c \psi \pi}$ can be evaluated by using Eq.~(\ref{Eq:PW}). As for $Z_{cs}(3985)$, the experimental measurements are still not abundant, here, we assume that the branching ratio of $Z_{cs}(3985) \to J/\psi K$ is the same as the one of $Z_c(3900) \to J/\psi \pi$ by considering the $\mathrm{SU}(3)$ symmetry. As for $P_c$ states, only the $J/\psi p$ decay mode have been observed experimentally at present.

To date, The pentaquark states $P_c$ were only observed in the $J/\psi p$ channel, which indicates that the $J/\psi p$ should be one of important decay modes of $P_c$ states. The branching ratios of $P_c \to J/\psi p$ were estimated from several percent to several tens percent depending one different model~\cite{Wu:2019rog}. In the present work, the branching ratios of $P_c \to J/\psi p$ are assumed to be $10\%$. The the coupling constants $g_{P_c \psi p}$ can be evaluated by using Eq.~(\ref{Eq:PW}) and the measured widths of $P_c$ states. All the estimated coupling constants have been collected in Table~\ref{Tab:CP}.

\begin{figure}[htb]
\includegraphics[width=8.0cm]{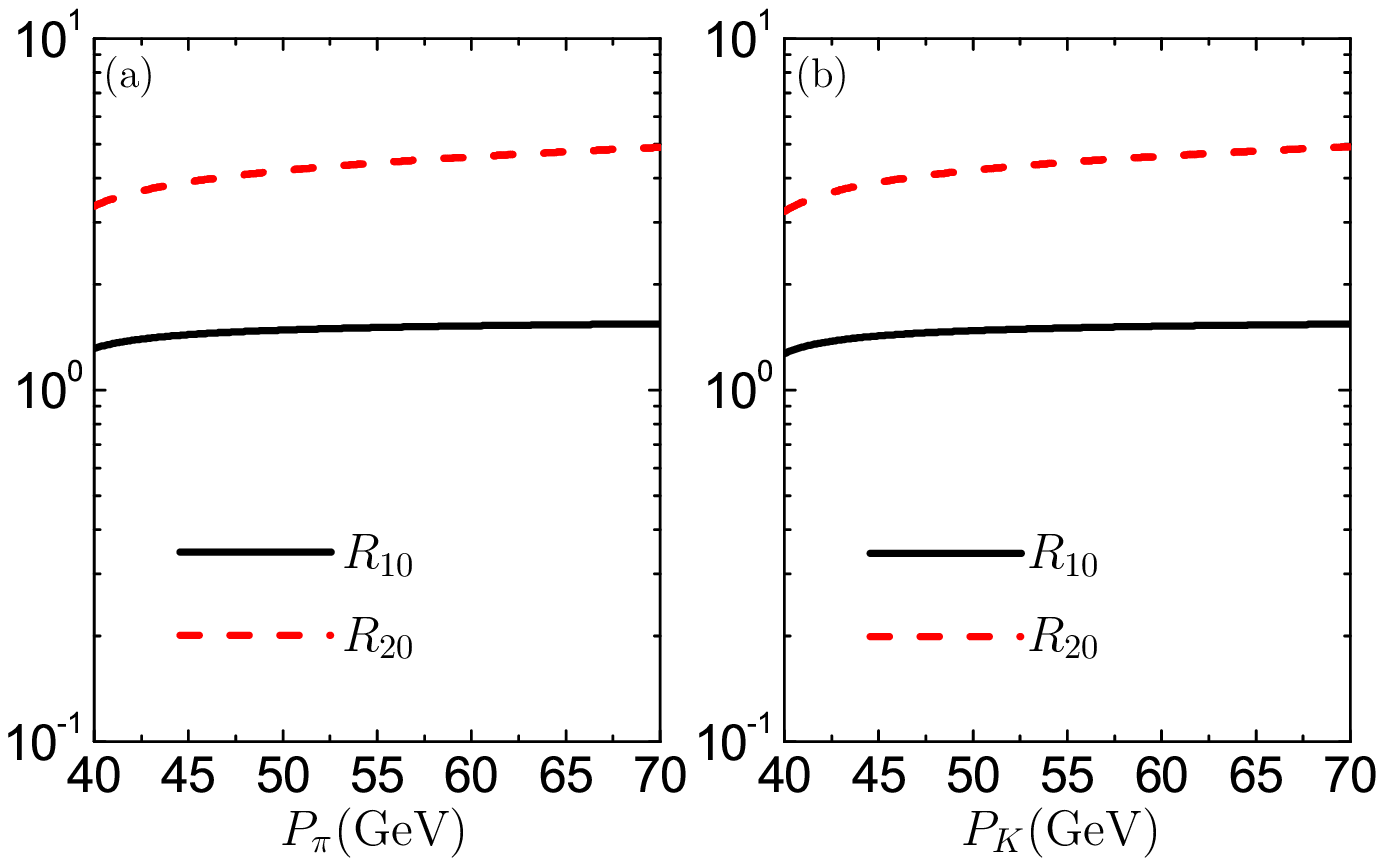}
\includegraphics[width=8.0cm]{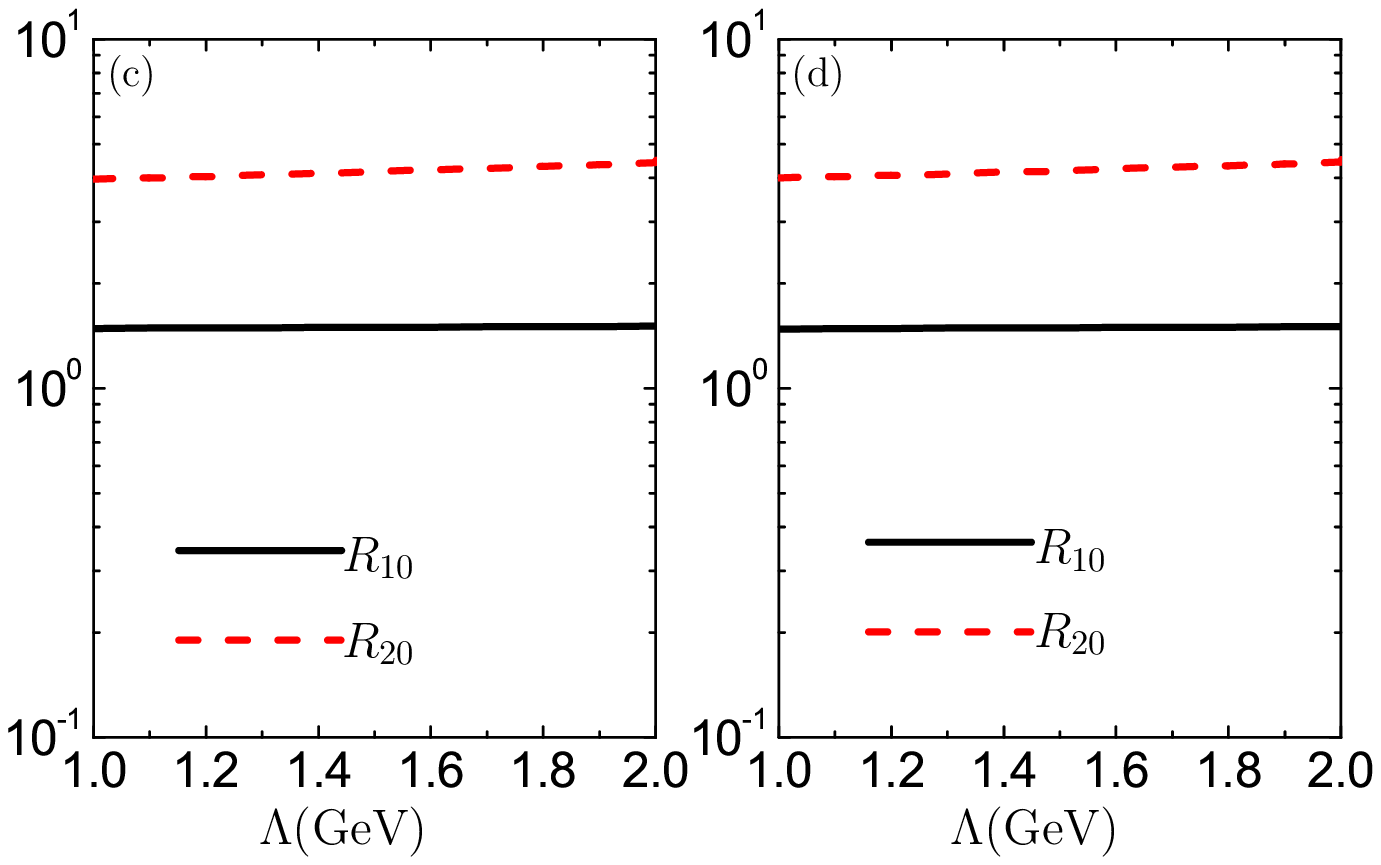}
 \caption{The cross sections ratios depending on beam energy and model parameter $\alpha$ . Diagram $(a)$ and $(b)$ present the beam energy dependences of the ratios of the cross section for $\pi p \to Z_c(3900) P_c$ and $K p \to Z_{cs}(3985) P_c$ with $\Lambda=2.0$, respectively. Diagram $(c)$ and $(d)$ present the model parameter dependences of the ratios   of the cross section for $\pi p$ and $Kp$ reaction with beam energy to be 55 GeV, respectively. \label{Fig:ratio} }
\end{figure}

\subsection{Cross sections for double exotic states production processes}
With the above preparation, we can estimate the cross sections for the double exotic states production processes. The cross sections for $\pi p \to Z_c(3900) P_c$ are presented in Fig.~\ref{Fig:CS-pip}, where diagrams (a), (b) and (c) correspond to $P_c(4312)$, $P_c(4440)$ and $P_c(4457)$ productions, respectively. For comparison, we take the same model parameter $\Lambda$ as the one used in the estimation of $\pi p \to Z_c(3900) p$~\cite{Huang:2015xud}, and then we can compare the cross sections for double exotic production process with the one of $\pi p \to Z_c(3900) p$~\cite{Huang:2015xud}. The cross sections for $\pi p \to Z_c(3900) P_c$ increase very fast near the thresholds and then become rather flat at higher pion momentum region. The cross sections can be greater than $10 $ nb when the momentum of pion is 70 GeV when $\Lambda=1.0$ GeV. In particular, when including the effect of model parameter, we find the cross sections for $\pi p \to Z_c(3900) P_c(4312)$ is $2 \sim 9 $  nb when $\Lambda$ varying from $1.0$ GeV to $2.0$ GeV at $P_\pi=70$ GeV. As for $\pi p \to Z_c(3900) P_c(4440)$ and $\pi p \to Z_c(3900) P_c(4457)$, the cross sections at $P_{\pi}=70$ GeV are $3\sim 13$ nb and $11 \sim 37$ nb, respectively.  The cross sections for $K p \to Z_{cs}(3985) P_{c}$ are presented in Fig.~\ref{Fig:CS-Kp}, where diagrams (a), (b) and (c) correspond to to $P_c(4312)$, $P_c(4440)$ and $P_c(4457)$ productions, respectively. The behaviors of the cross sections are very similar to those of $\pi p \to Z_{c}(3900)P_c$, which show sharp increasing near the threshold but the cross sections for $Kp\to Z_{cs}(3985) P_c$ are about $2.5$ times smaller than those for $\pi p \to Z_c(3900) P_c$ due to the coupling constants $g_{Z_{cs} \psi K}$ is smaller than the one of $g_{Z_{c} \psi \pi}$.

\begin{figure*}[h!]
\includegraphics[width=16cm]{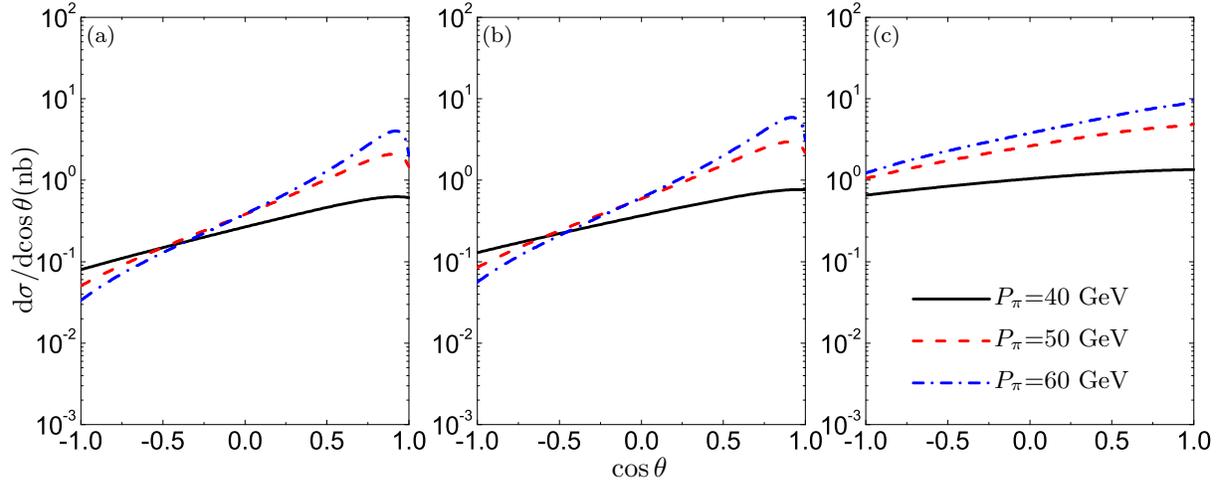}
 \caption{The same as Fig.~\ref{Fig:CS-pip} but for differential cross sections depending on $\cos \theta$ \label{Fig:dCS-pip} }
\end{figure*}

\begin{figure*}[h!]
\includegraphics[width=16cm]{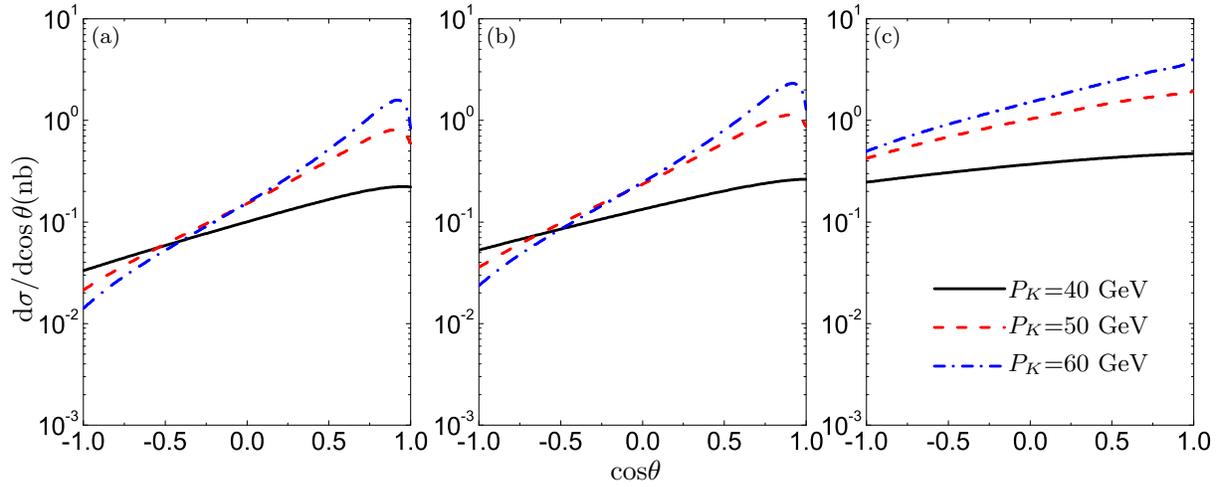}
 \caption{The same as Fig.~\ref{Fig:dCS-pip} but for  $Kp\to Z_{cs}(3985) P_c$ processes, where diagrams (a), (b) and (c) correspond to $P_c(4312)$, $P_c(4440)$ and $P_c(4457)$ production, respectively. \label{Fig:dCS-kp} }
\end{figure*}

As shown in Figs.~\ref{Fig:CS-pip}-\ref{Fig:CS-Kp}, the energy and model parameter dependences of the cross sections for $\pi p \to Z_c(3900) P_c$ and $K p \to Z_{cs} P_c$ for $P_c(4312)$, $P_c(4440)$ and $P_c(4457)$ are very similar, then one can define cross sections ratio as following,
\begin{eqnarray}
R_{10}=\frac{\sigma(\pi p\rightarrow Z_{c}(3900)P_{c}(4440))}
       {\sigma(\pi p\rightarrow Z_{c}(3900)P_{c}(4312))},\nonumber\\
R_{20}=\frac{\sigma(\pi p\rightarrow Z_{c}(3900)P_{c}(4457))}
       {\sigma(\pi p\rightarrow Z_{c}(3900)P_{c}(4312))},
\end{eqnarray}
which are expected to be independent on the energy and the model parameter. In Fig.~\ref{Fig:ratio}, the beam energy and model parameter dependences of the ratios of the cross sections for $\pi p \to Z_c(3900) P_c$ and $K p \to Z_{cs}(3985) P_c$ are presented. As shown in Fig.~\ref{Fig:ratio}-(c) and (d), the ratios are entirely independent on the model parameter $\Lambda$. Our estimations indicate the cross sections for $P_c(4457)$ productions from $\pi p$ and $Kp$ reactions are about 3 times larger than those of $P_c(4440)$ and $P_c(4312)$, while cross sections for $P_c(4440)$ and $P_c(4312)$ are almost the same.

Besides the cross sections, the differential cross sections depending on scattering angle related to the outgoing $P_c$ and the incident pion/kaon beam are also estimated. In Fig.~\ref{Fig:dCS-pip}, we present the differential cross section for $\pi p \to Z_c(3900) P_c$ depending on $\cos \theta$, where diagrams (a), (b) and (c) correspond to $P_c(4312)$, $P_c(4440)$ and $P_c(4457)$ production, respectively. Here, the pion beam momentum are set to be 40, 50 and 60 GeV, respectively. Our estimation indicate that the differential cross sections reach their maximum at the forward angle limit, i.e., $\cos\theta=1$. For the case of $P_\pi=40$ GeV, the beam energy is just a bit larger than the threshold of the reaction, thus the differential cross sections do not strongly depend on the scattering angle. With the $P_\pi$ increasing, more produced pentaquark or tetraquark states are concentrated in the forward angle area. Moreover, the $\cos\theta$ dependence of the differential cross sections are very similar for $P_c(4312)$, $P_c(4440)$ and $P_c(4457)$ due to the similarity of these double exotic production process.

  The differential cross sections for $Kp\to Z_{cs}(3985)P_c$ are presented in Fig.~\ref{Fig:dCS-kp}, where the kaon beam momentum are also set to be 40, 50 and 60 GeV, respectively. Our estimations indicate that the scattering angle dependences of the cross sections are almost the same as those of $\pi p \to Z_c(3900) P_c$ due to the similarity between these two kinds of double exotic production processes.

\section{Summary}
\label{Sec:Summary}
As the typical tetraquark and pentaquark candidates, $Z_c/Z_{cs}$ and $P_c/P_{cs}$ have stimulated theorists' great interest and after the observations their properties have been investigated from various aspects, such as mass spectra, decay behaviors and production processes, which are expected to reveal the inner structure of these multiquark states.

 In the present work, we evaluate the experimental potential of the double exotic production in the $\pi p$ and $Kp$ scattering process. The cross sections for $\pi p \to Z_{c}(3900) P_c$ and $Kp\to Z_{cs}(3985) P_c$ are evaluated. Our estimations indicate that the cross sections are dependent on the model parameter and the beam energy. In particular, at $P_\pi =70$ GeV the cross sections for $\pi p \to Z_c(3900) P_c$ can reach up to 9,  13 and 37 nb for $P_c(4312)$, $P_c(4440)$ and $P_c(4457)$ at $\Lambda$=1.0, respectively. Moreover, the beam energy and model parameter dependences of the cross sections are very similar for three $P_c$ states, thus the ratios of these cross section are almost independent on the beam energy and model parameter, which can be tested by further experimental measurements. In addition, the beam energy and model parameter dependences of the cross sections for $K p\to Z_{cs}(3985) P_c$ are almost the same as those of $\pi p \to Z_c(3900) P_c$. Besides the cross sections, the differential cross sections for the considered processes are also evaluated and our estimations indicate that more produced pentaquark or tetraquark states are concentrated in the forward angle area.

 At the end of the present work, we would like to discuss the experimental potential of these double exotic production processes by comparing the double exotic production process of $\pi p \to Z_{c}(3900) P_c$ with $\pi p \to Z_c(3900) p$~\cite{Huang:2015xud}. In Ref.~\cite{Huang:2015xud}, the cross section for $\pi p \to Z_c(3900) p$ at $P_\pi =70$ GeV is of order of  100 nb with the assumption that $J/\psi \pi$ should be the dominant decay channel of $Z_c(3900)$. However, the experimental measurements indicated that the branching fraction of $Z_c(3900)\to J/\psi \pi$ should be of order $10\%$ as indicated in Eq.~(\ref{Eq:BRZc}), thus the cross sections for $\pi p \to Z_c(3900) p$ should be about 10 nb at $P_\pi =70$ GeV, which is the same order as the cross sections for $\pi p\to Z_c(3900) P_c(4312)/P_c(4440)$ and about several times smaller than the one of $\pi p \to Z_c(3900) P_c(4457)$. Thus, if the experimental measurement in COMPASS or J-PARC could detect $Z_c(3900)$ in the $\pi p\to Z_c(3900) p$ process, the double exotic production process, especially $\pi p \to Z_c P_c(4457)$, discussed in the present work should be more easily accessible in these facilities.

\section*{Acknowledgement}
This work is supported in part by the National Natural Science Foundation of China (NSFC) under Grant Nos.11775050 and 11675228.

\end{document}